\documentclass[slac_one]{revtex4}
\pdfoutput=1
\usepackage{graphicx}
\usepackage{fancyhdr}
\newenvironment{packed_enum}{
\begin{itemize}
  \setlength{\itemsep}{1pt}
  \setlength{\parskip}{0pt}
  \setlength{\parsep}{0pt}
}{\end{itemize}}

\begin{document}

\title{Early b-physics at CMS} 

%

\author{Andrea Rizzi}
\affiliation{ETH Zurich, Switzerland}

\begin{abstract}
The CMS experiment at the Large Hadron Collider collected in the first months of operation a luminosity of about 300/nb. The first results in the context of the B physics obtained with these data are presented. The di-muon resonances from J/$\psi$ and $\Upsilon$ decays are presented and their total and differential cross-sections measured.\\
The inclusive B production have also being investigated and two independent measurements are reported. Muons in jets are used as a way to identify events with B content, the kinematic properties (ptRel) of the muons are used to separate the B production from other processes producing muons in jets. The second measurement of inclusive B production is done using b-tagging techniques and higher energy jets. The two measurements cover different phase spaces, comparison with LO and NLO prediction are also presented.
\end{abstract}
\maketitle

\thispagestyle{fancy}

\section{The CMS detector}
A detailed description of the Compact Muon Solenoid (CMS) experiment can be found elsewhere\cite{cms}. The central feature of the CMS apparatus is a superconducting solenoid, of 6 m internal diameter. Within the field volume are the silicon tracker, the crystal electromagnetic calorimeter (ECAL) and the brass/scintillator hadron calorimeter (HCAL). Muons are detected
in the pseudorapidity window $|\eta|< 2.4$, by gaseous detectors made of three technologies: Drift
Tubes (DT), Cathode Strip Chambers (CSC), and Resistive Plate Chambers (RPC), embedded
in the iron return yoke. \\
The silicon tracker is composed of pixel detectors (three barrel layers and two forward disks in either side of the detector, made of 66 million $100 \times 150 \mu m^2$ pixels) followed by microstrip detectors (several layers with strips of pitch between 80 and 180 $\mu m$). Thanks to the strong magnetic field, 3.8 T, and to the high granularity of the silicon tracker, the transverse momentum, $p_T$ , of the muons matched to reconstructed tracks is measured with a resolution
better than ∼ 1.5\% for $p_T$ smaller than 100 GeV. 
\subsection{Performance of tracking detectors}
\begin{figure}
\includegraphics[width=7.5cm,height=5.3cm]{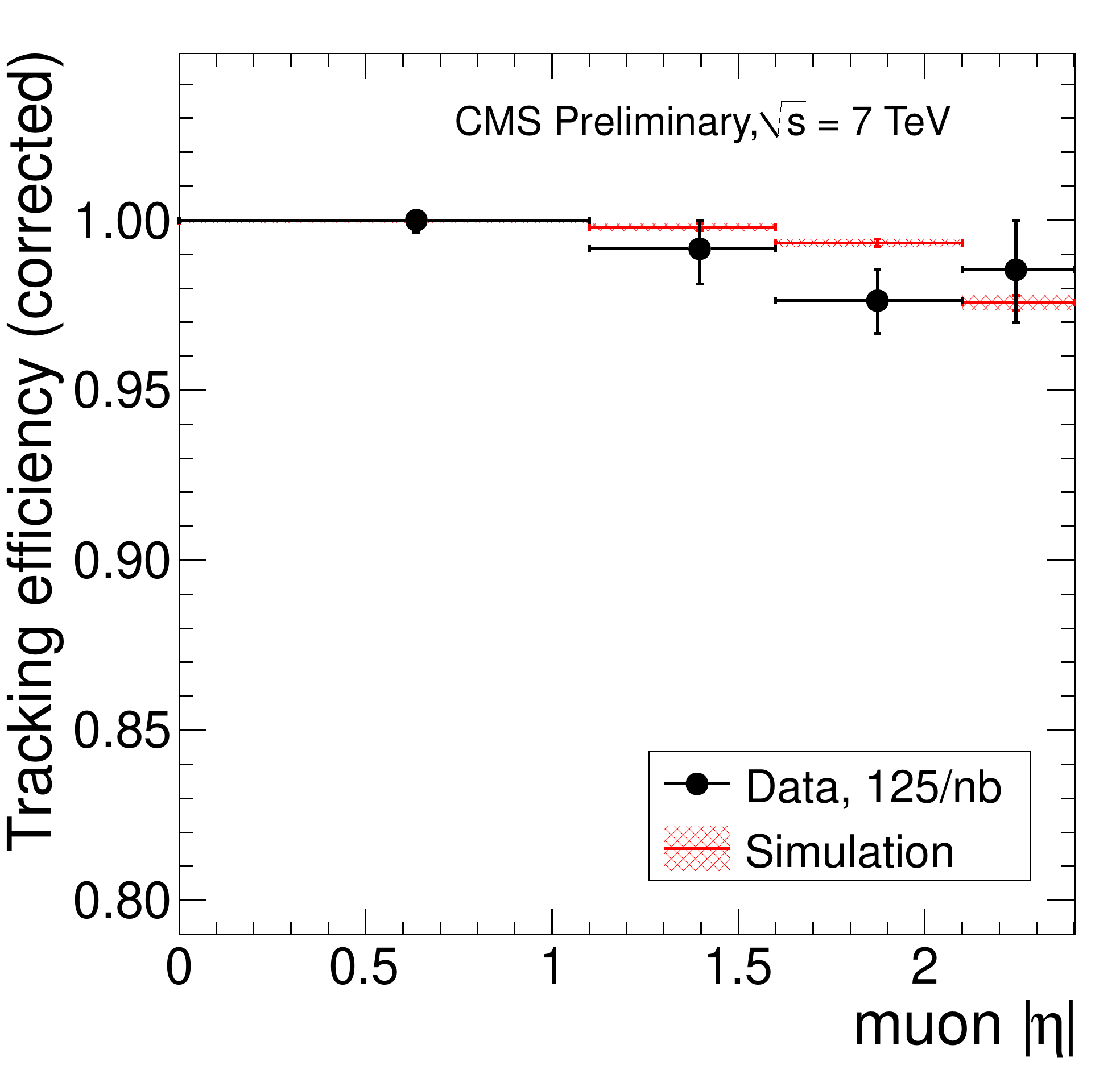}
\includegraphics[width=7.5cm,height=5.3cm]{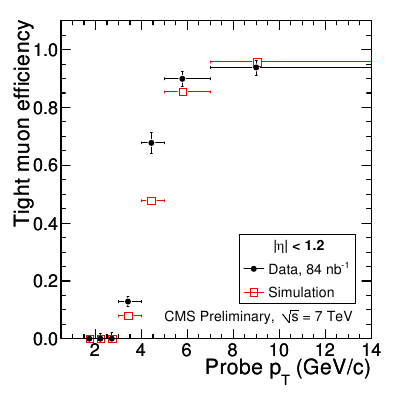}
\caption{Tracking efficiency from tag and probe for inner tracking (left) and outer muon reconstruction (right) as a function of the pseudorapidity $\eta$.}
\label{fig:eff} 
\end{figure}

\begin{figure}
\includegraphics[width=7.5cm,height=5.3cm]{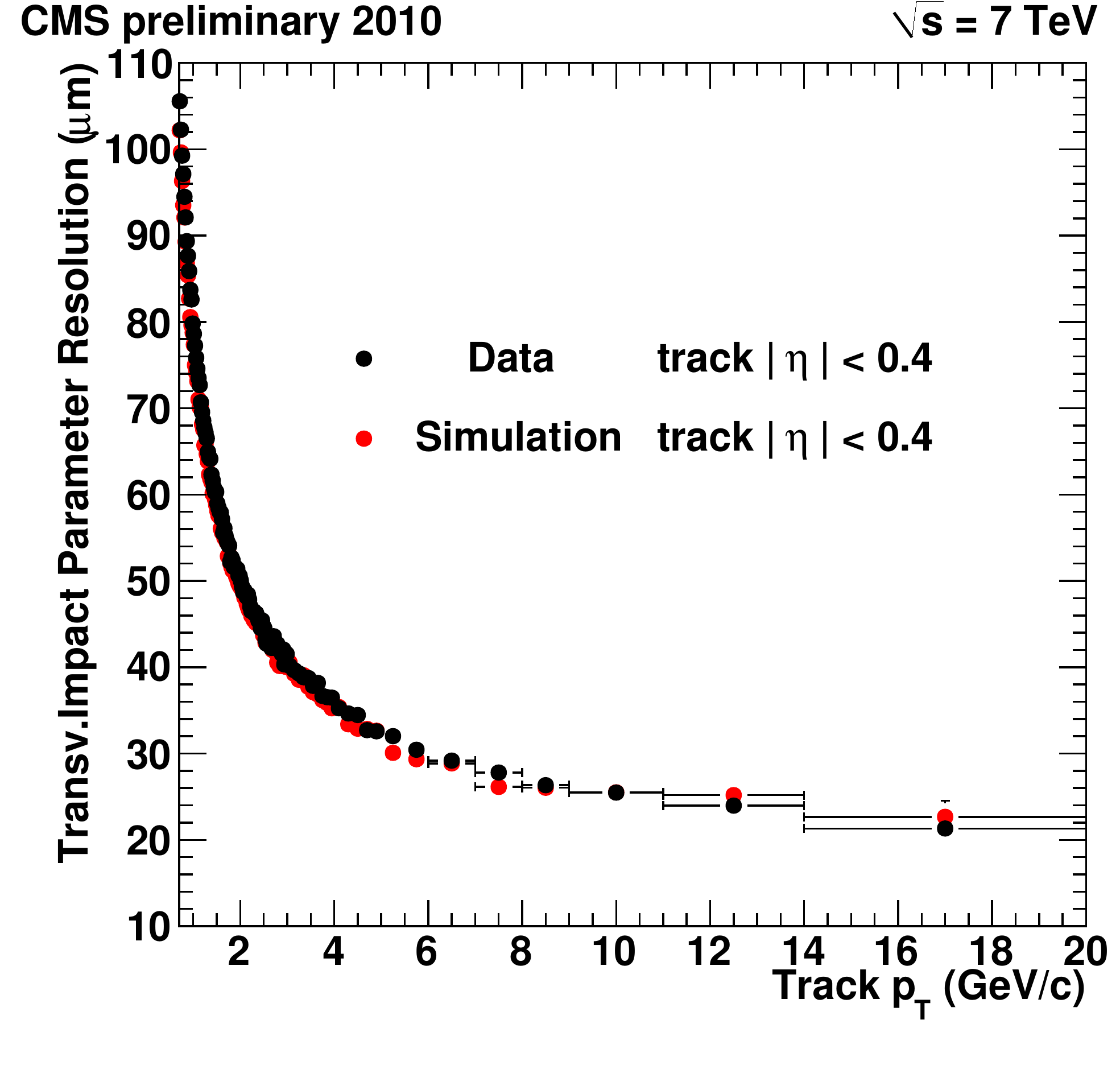}
\includegraphics[width=7.5cm,height=5.3cm]{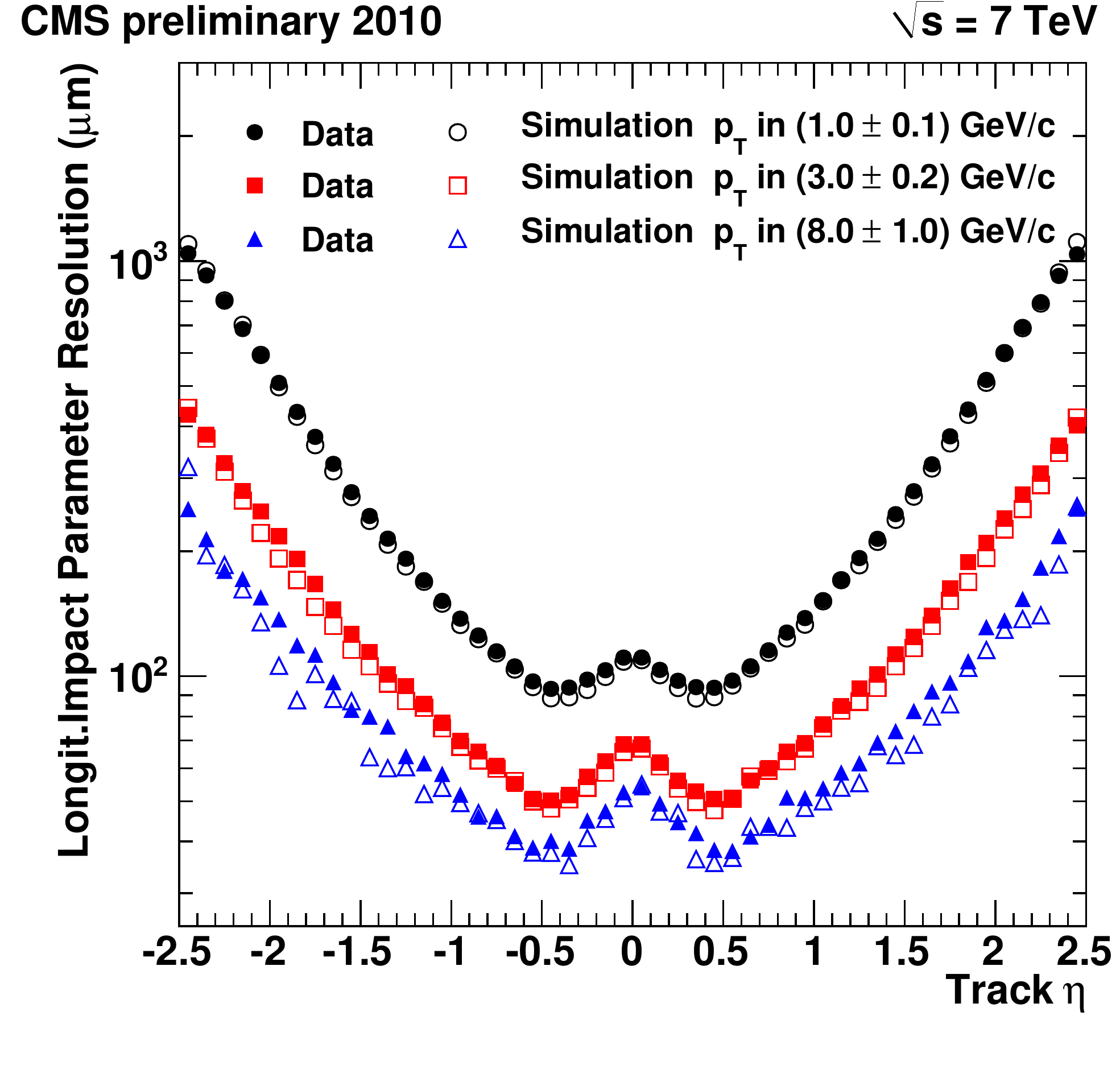}
\caption{Resolution for transverse impact parameter as a function of transverse momentum (left) and longitudinal impact parameter as a function of pseudorapidity (right).}
\label{fig:ip}  
\end{figure}

\begin{figure}
\includegraphics[width=7.5cm,height=5.3cm]{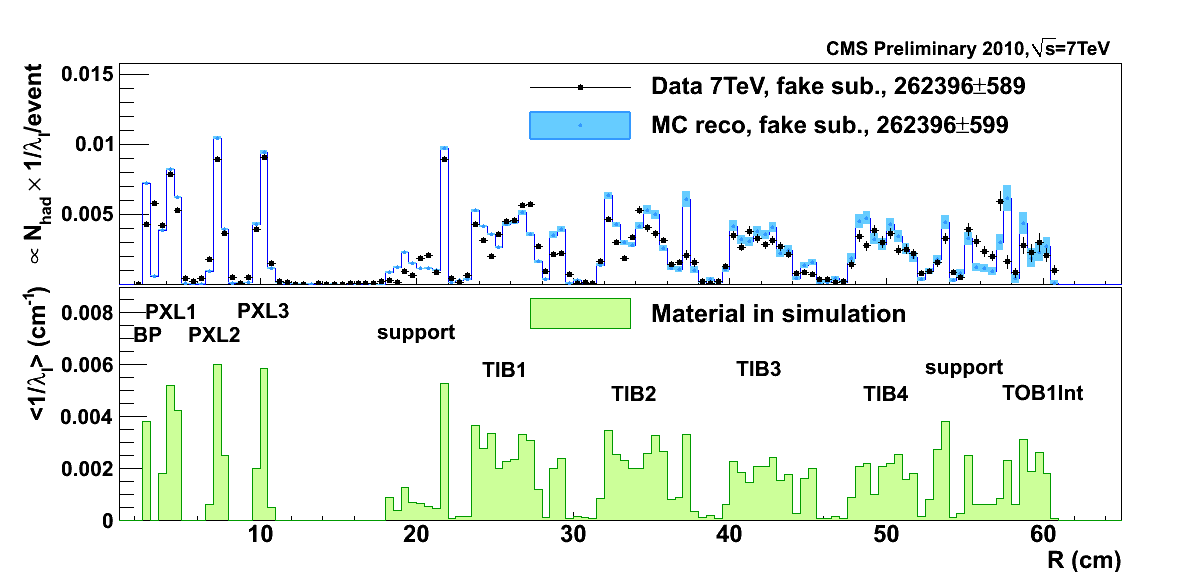}
\includegraphics[width=7.5cm,height=5.3cm]{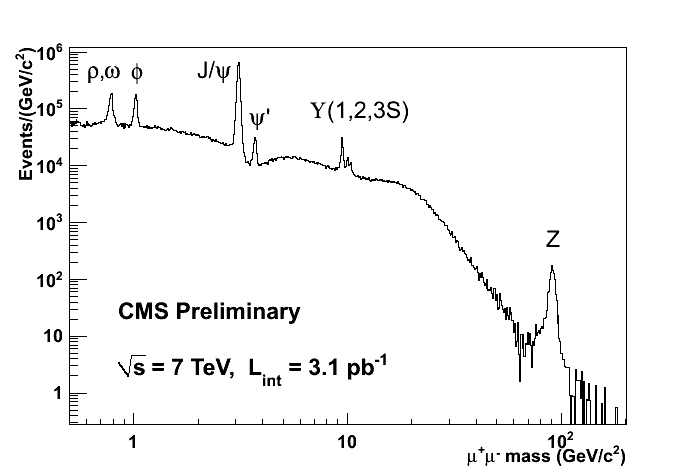}
\caption{Nuclear interaction distribution as a function of the transverse radius (left). Di-muon mass spectrum (right). }
\label{fig:material_and_spectrum}   
\end{figure}

The key element for the b-physics program of CMS are the tracking detectors, both the outer muon spectrometer and the inner silicon tracker. The performance of those detectors and readiness of the software used to interpret their data have been measured and tested with first LHC data at 7~TeV.\\
The tracking efficiency has been measured in different ways as documented in \cite{TRK2}\cite{MUO2}. The so called \emph{tag and probe} approach allows to study both the efficiency of the inner tracker and the one of the muon spectrometer in triggering, reconstructing and selecting a muon track in the transverse momentum range typical of J/$\psi$ production. 
This method consists in reconstructing the J/$\psi$ resonance using partial information from one of the two muons (e.g. only the outer muon spectrometer information) and then counting how often the unused information (e.g. the inner tracker track) is actually reconstructed \cite{MUO2}\cite{BPH2}. The results for inner tracker and outer muon efficiency are shown in Figure \ref{fig:eff}.\\
The CMS pixel detector allows a precise measurement of the track impact parameter, this is needed in order to efficiently apply \emph{b} jet identification techniques (b-tagging). The resolution on the impact parameter has been measured using the first LHC data \cite{TRK5} by looking at the core of the distribution of the impact parameter with respect to the reconstructed primary vertex after subtracting the vertex resolution. The results are shown in Figure \ref{fig:ip}.
In addition to the above direct measurements, the reliability of the simulation software has been checked by verifying the track momentum scale and the amount of material crossed by a particle produced at the interaction point while flying through the tracker detector\cite{TRK3}\cite{TRK4}. The latter, especially for what concern the innermost detector, is an important element in order to correctly reproduce in the simulation the impact parameter resolution observed in data. An accurate description of the materials has been implemented in the CMS software. The material is studied by reconstructing the position of nuclear interactions and gamma conversions and comparing their distributions with those predicted by the simulation. The results are shown in the left plot of Figure \ref{fig:material_and_spectrum}. An agreement at the level of about $10\%$ is observed.
\section{Di-Muon resonances}
The muons are reconstructed with two different algorithms (\emph{global muons} and \emph{tracker muons}). The first algorithm (global) uses the information of all detectors both for identifying and for fitting the muon track property. The second algorithm (tracker) does not use the outer muon spectrometer information in the fitting procedure. The requirement for a \emph{tracker muon} are hence looser than for a \emph{global muon} and the algorithm is expected to be more efficient at low $p_T$. \\

The following selection is applied to muons in events triggered by low $p_T$ muons:
\begin{packed_enum}
 \item at least 12 hits in the tracker detector (at least 2 of them in the pixels detector)
 \item $\chi^2$/number of degree of freedom $< 4$
 \item transverse impact parameter $< 3$~cm
 \item longitudinal distance from the primary vertex  $< 30$~cm
\end{packed_enum}
The events are then analyzed if at least two muons are passing the selection.\\
The resulting spectrum for the invariant mass of the two muons is shown in Figure \ref{fig:material_and_spectrum}: the known di-muon resonances are well visible. In the following paragraph the analysis performed in the J/$psi$ and $\Upsilon$ regions are described.
\subsection{J/$\psi$ and $\Upsilon$}
\begin{figure}
\includegraphics[width=7.5cm,height=5cm]{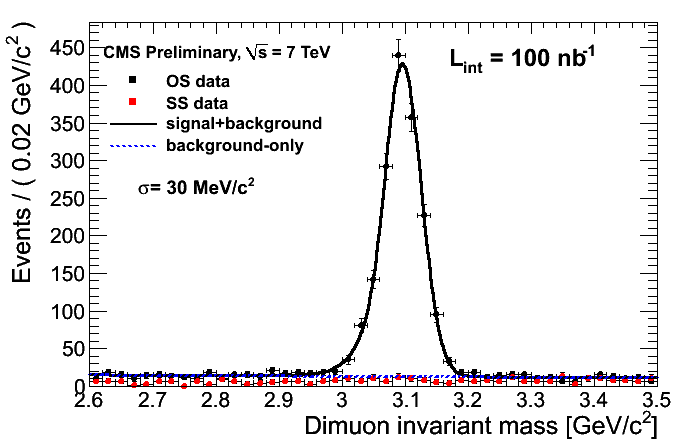}
\includegraphics[width=7.5cm,height=5cm]{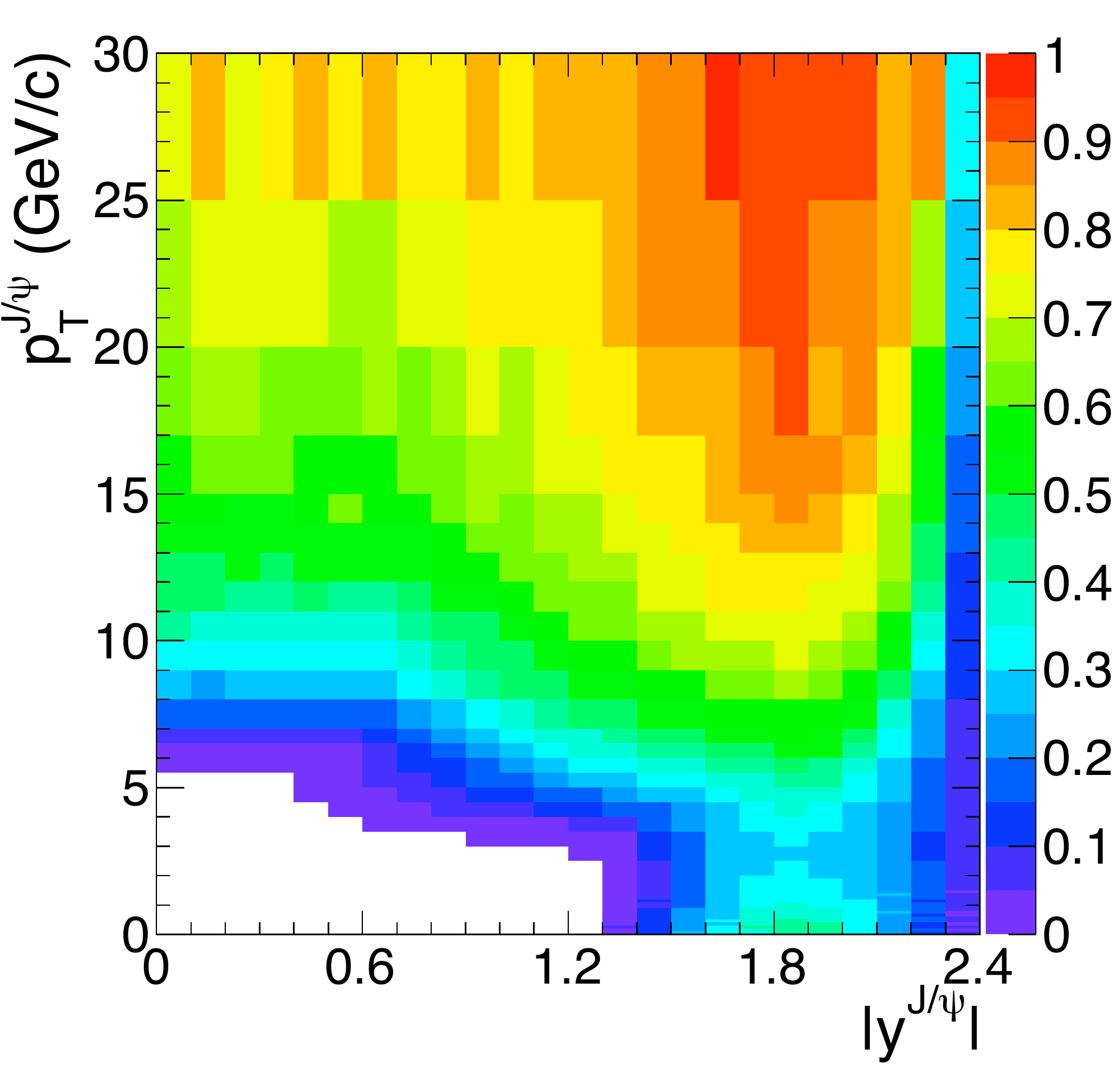}
\caption{Di-muon spectrum near the $J/\psi$ mass (left). Acceptance map as a function of $J/\psi$ momentum and rapidity (right).}
\label{fig:jpsi} 
\end{figure}

\begin{figure}
\includegraphics[width=7.5cm,height=5cm]{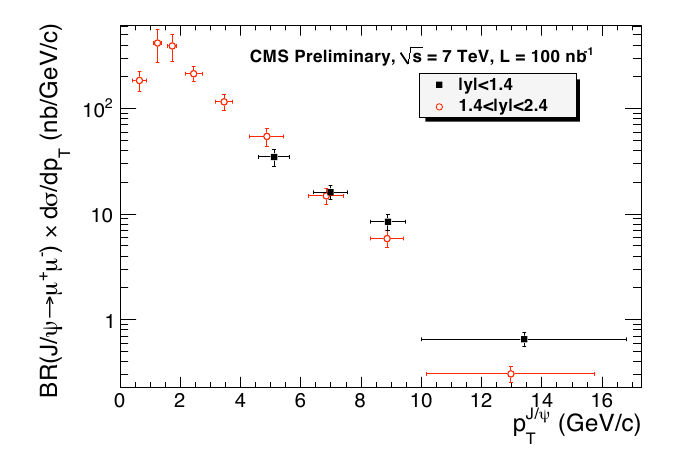}
\includegraphics[width=7.5cm,height=5cm]{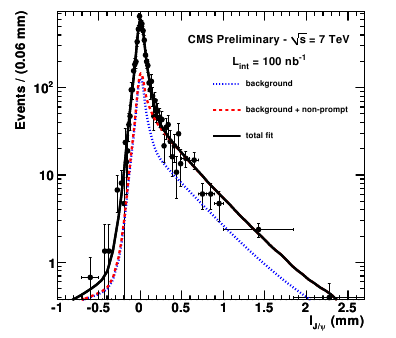}
\caption{Transverse momentum differential cross section for $J/\psi \rightarrow \mu \mu$ (left). Lifetime fit to extract non-prompt $J/\psi$ fraction (right). } 
\label{fig:jpsi2} 
\end{figure}

\begin{figure}
\includegraphics[width=7.5cm,height=5cm]{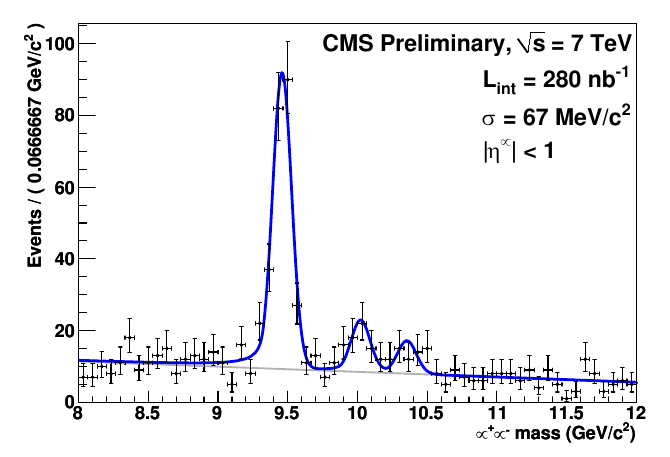}
\includegraphics[width=7.5cm,height=5cm]{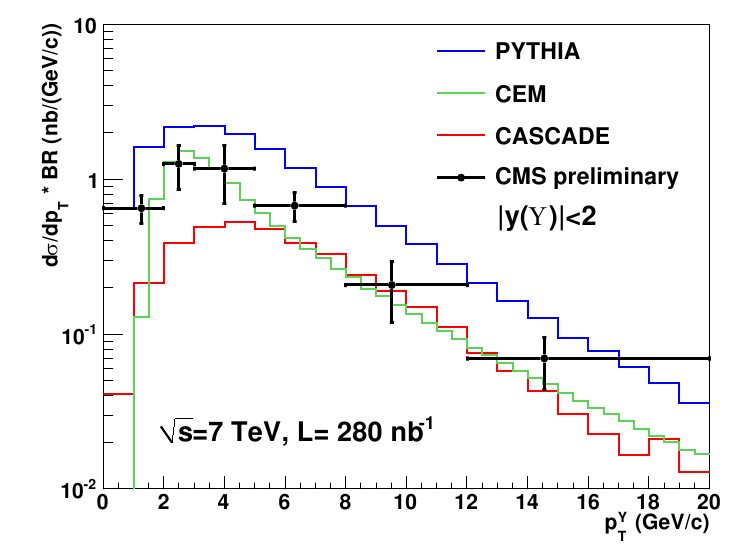}
\caption{Mass spectrum near the $\upsilon$ resonance (left). Transverse momentum differential cross section (right).}
\label{fig:upsilon} 
\end{figure}

To select the events with $J/\psi$ decays, muons with opposite charge are paired and their invariant
mass is computed. The invariant mass of the muon pair is required to be between 2.6 and
3.5 GeV . The two muons helices are fitted with a common vertex constraint, and events
are retained if the fit $\chi^2$ probability is larger than 0.1\%. Pairs made of different muon type
combinations are reconstructed: two global muons, two tracker muons or one global and one
tracker muon. In case of multiple combinations in the same event, the combination with the
purest muon content is chosen (global muon being purer than tracker muons). If two candidate pairs belong to the same dimuon type combination, the one with the largest pT is chosen. On
average 1.07 J/$\psi$ combinations were found per event. The analysis is then performed summing
the three categories.
Same sign dimuons are also reconstructed, and are used as a check of the background level. The mass peak is shown in Figure \ref{fig:jpsi}.\\
The observed number of J/$\psi$ events is corrected for the detector acceptance and muon reconstruction efficiency. The acceptance accounts for purely geometrical and kinematical limitations
and is taken from the simulation, while the efficiency is related to instrumental effects which
can be measured from data. The acceptance map is shown in Figure \ref{fig:jpsi}. The acceptance is a function of the kinematic properties of the decay products, hence it depends on the J/$\psi$ polarization. The measurement of the polarization is not discussed here, the results are given for the unpolarized case and in \cite{BPH2} for other polarization scenarios.\\
The  J/$\psi$ yields are measured in bins of rapidity and transverse momentum, the resulting efficiency-corrected differential cross section is shown in the left plot of Figure \ref{fig:jpsi2}. The fraction of  J/$\psi$ from B decay is extracted with a lifetime fit as shown in the right plot of Figure \ref{fig:jpsi2}.
A similar selection to the one used for J/$\psi$ is done for an invariant mass window near the $\Upsilon$ resonance ($8 < m_{\mu\mu} < 12$~GeV). The resulting mass spectrum and the differential cross section measured after the efficiency-correction are shown in Figure \ref{fig:upsilon}.

\section{Inclusive b-production cross section}
The inclusive b-production is investigated with two complementary approaches. The first exploits the decay of B to muons and the high mass of the B, resulting in muons having a large momentum component orthogonal to the B flight direction ($p_T^{rel}$). The second uses jets and b-tagging.
\subsection{$b \rightarrow \mu + X$}
\begin{figure}
\includegraphics[angle=90,width=5.3cm,height=5.1cm]{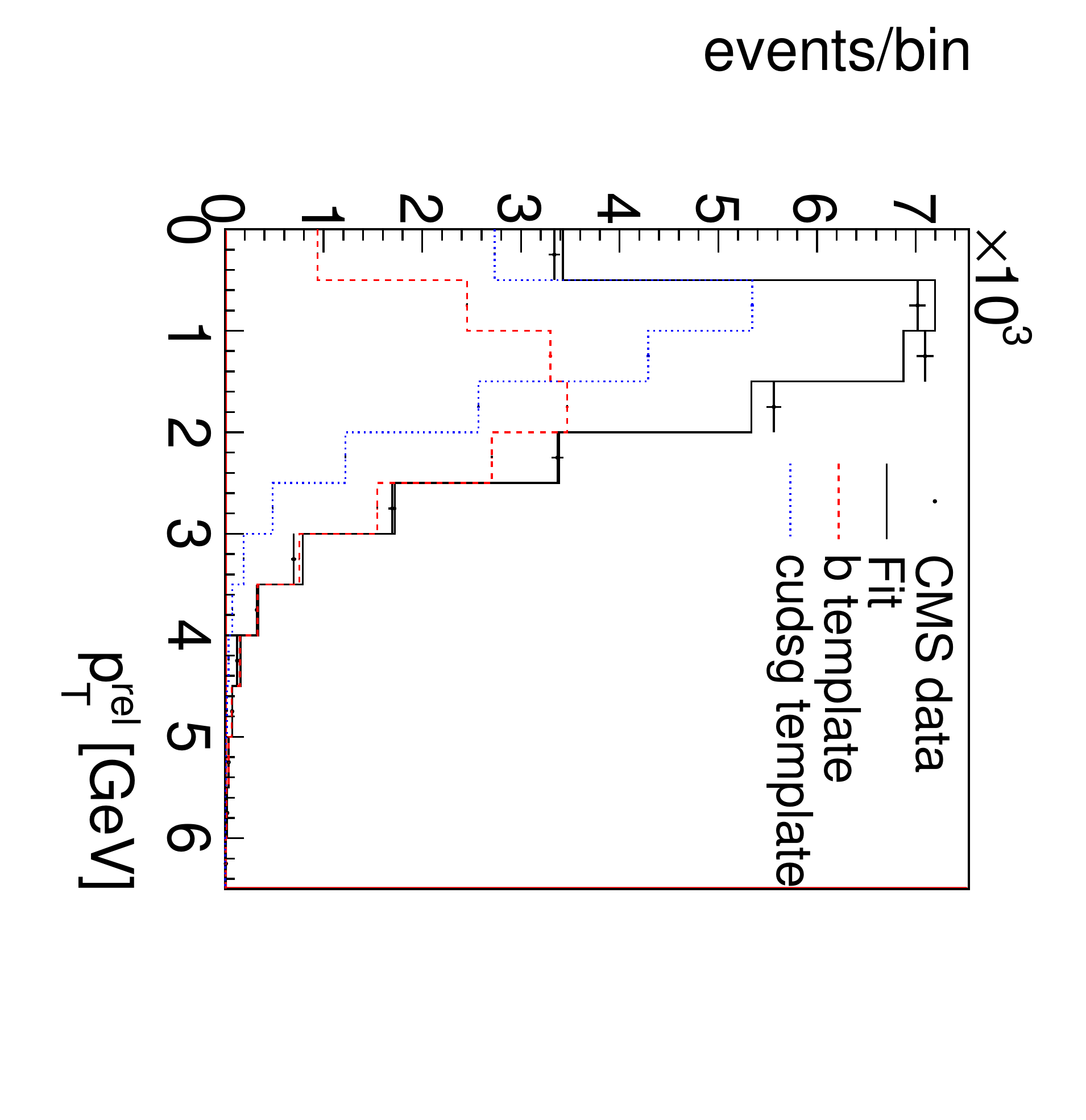}
\includegraphics[angle=90,width=5.7cm,height=5.4cm]{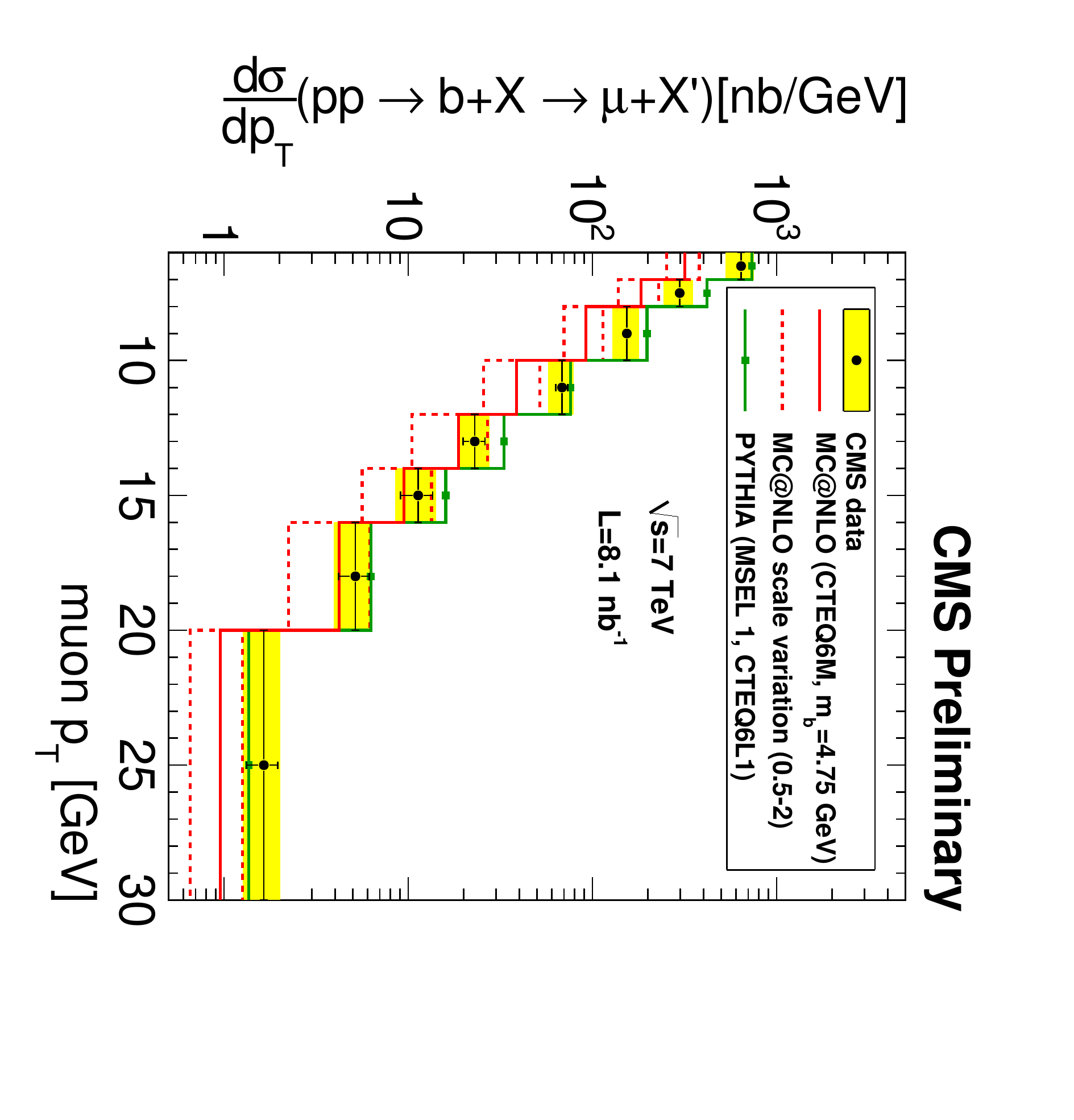}
\includegraphics[angle=90,width=6cm,height=5.6cm]{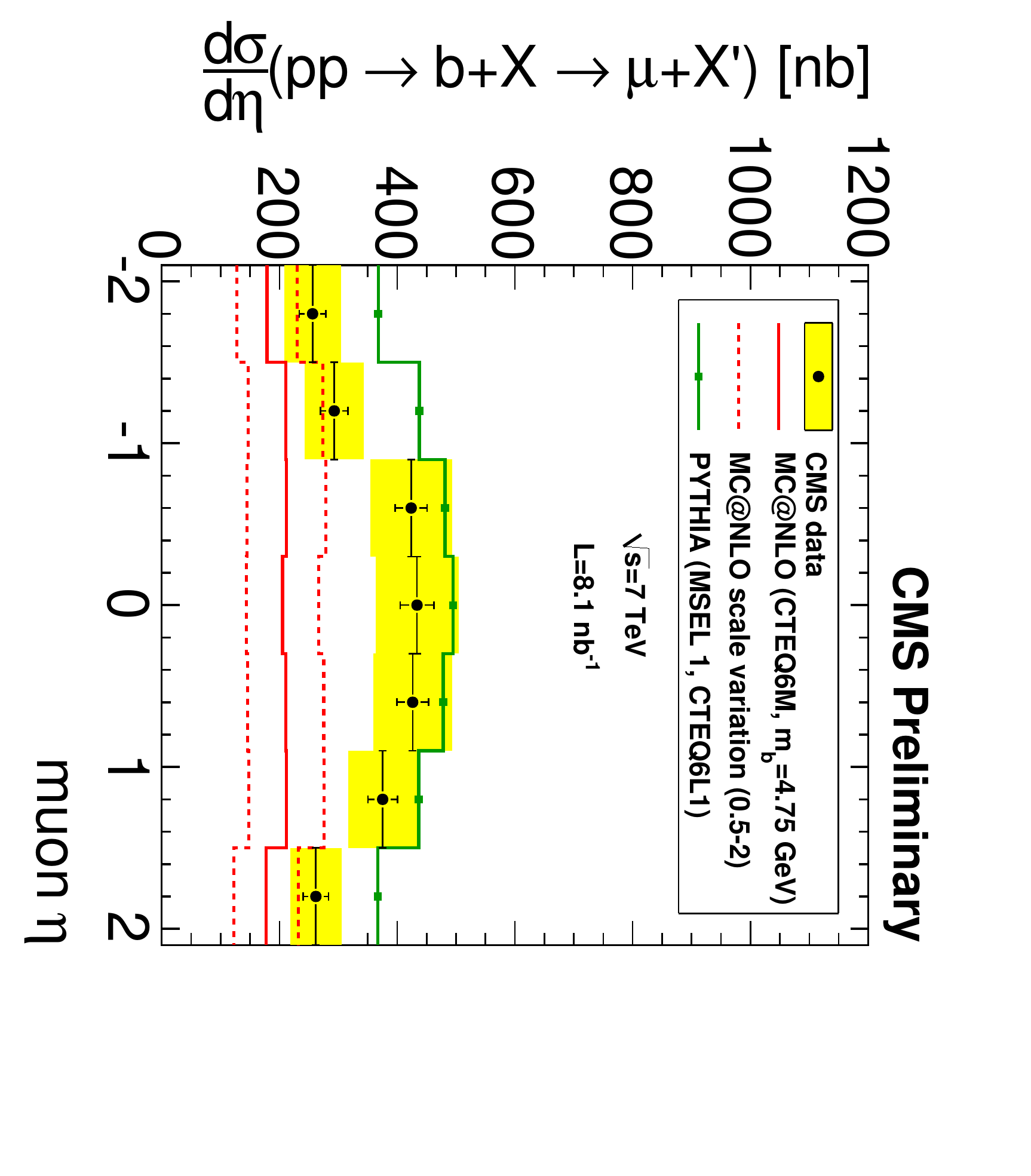}
\caption{$p_T^{rel}$ templates fit (left). Transverse momentum (center) and pseudorapidity (right) differential cross sections.}
\label{fig:ptrel} 
\end{figure}

The events are selected online with a single muon trigger with no explicit momentum cuts. In the data analysis the reconstructed muons are required to pass the following criteria: $p_T > 6$~GeV, $\eta < 2.1$, $\chi^2$/ndof~$< 10$, 12 hits in the tracker (2 in the pixels), longitudinal distance from the interaction point $< 20$~cm.\\
Track jets are made using the anti-kT algorithm on all tracks in the event with $p_T > 0.3$~GeV. The jets with $E_T> 1$~GeV are then considered.\\
The $p_T^{rel}$ is computed using the jet direction, excluding the muon track, as B flight direction hypothesis. The $p_T^{rel}$ is then fitted with two different templates for B and non-B contributions in different rapidity and transverse momentum bins. An example of fit is shown in the left plot of Figure \ref{fig:ptrel}.\\
After an efficiency-correction the differential cross section in $\eta$ and $p_T$ is obtained and compared with theoretical predictions as shown in the center and right plots of Figure \ref{fig:ptrel}.
\subsection{b-tagged jets}
\begin{figure}
\includegraphics[width=7cm]{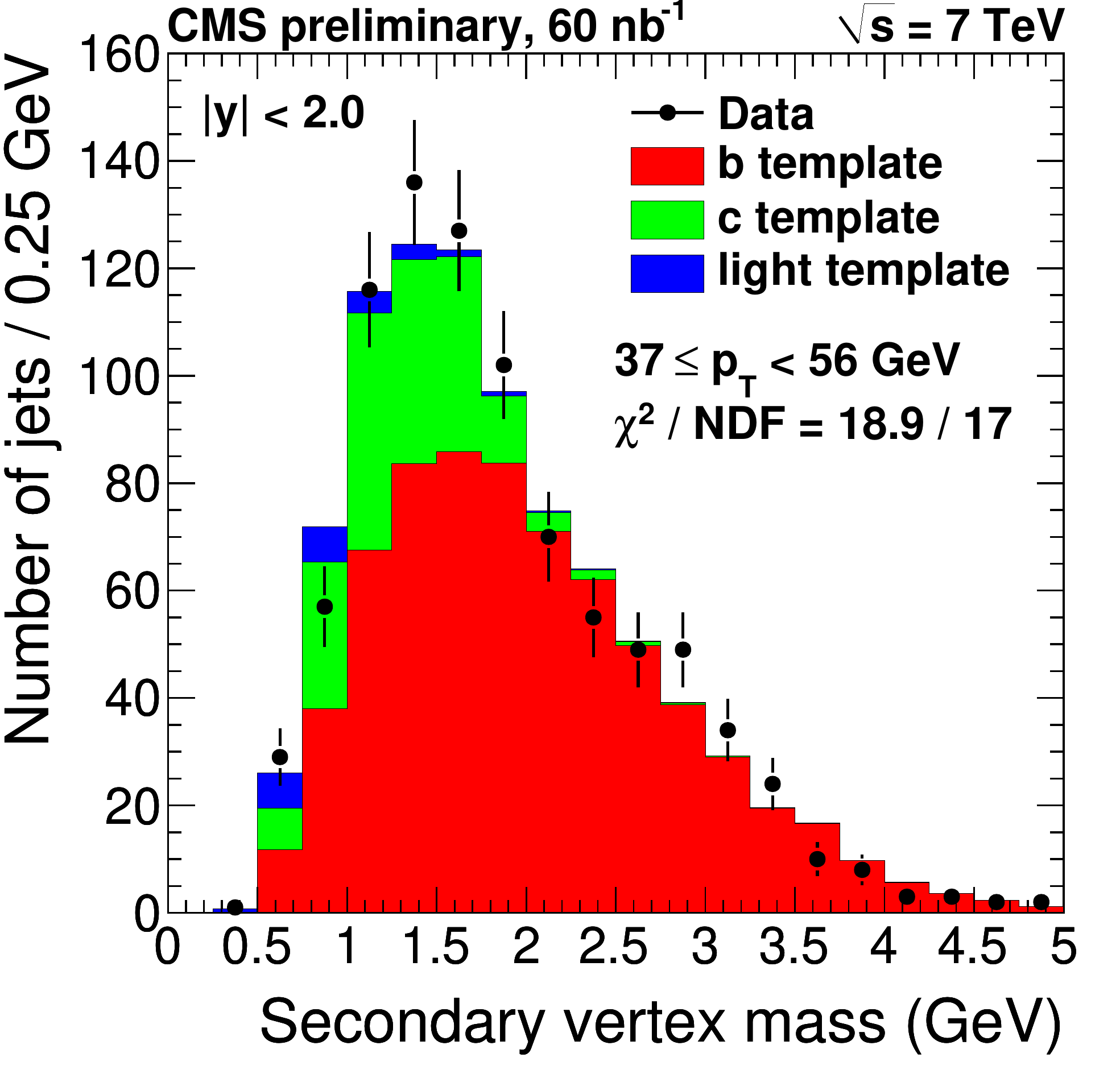}
\includegraphics[width=7cm]{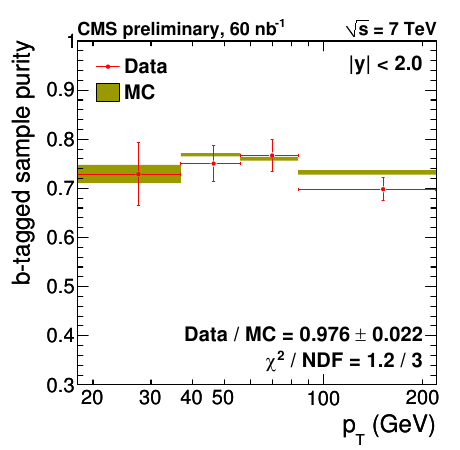}
\caption{Template fit on the vertex mass distribution (left) used to estimate the purity after b-tagging. The right plot shows, as a function of the $p_T$ the comparison of the purity derived from the fit and the one predicted in MC simulation.}
\label{fig:purity}
\end{figure}

An inclusive b-jet production measurement is obtained by applying b-tagging to the inclusive jet sample.\\
The inclusive jet data is collected using a combination of Minimum Bias and
single jet triggers, which are consecutively used in the lowest $p_T$ range
where the triggers are fully efficient. The raw $p_T$ spectra are unfolded using the ansatz
method~\cite{ansatz-one,ansatz-two}, with the jet $p_T$ resolution obtained
from MC.\\
The b jets are tagged using secondary vertices made of at least three charged tracks.
The b-tagging efficiency and the mistag rates from c-jet and light jet flavors
are taken from the MC simulation cross checked on data \cite{BTV-10-001}.\\
An additional check of the purity of the sample is obtained by fitting the secondary vertex invariant mass distribution with MC derived template functions. The purity obtained from MC simulation and from the template fit are compatible within the statistical uncertainty (Figure \ref{fig:purity}).

\begin{figure}
\includegraphics[width=7cm]{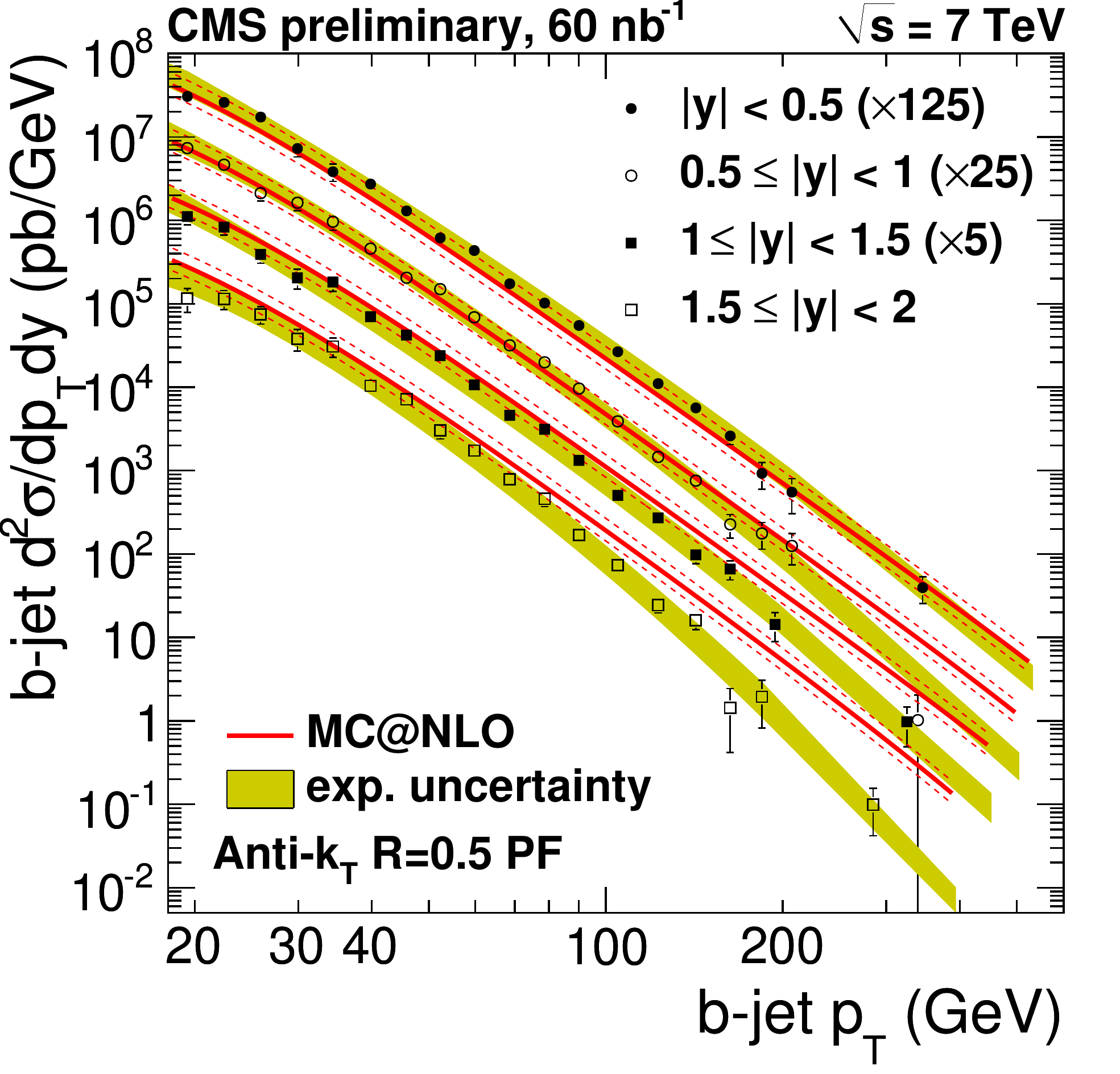}
\includegraphics[width=7cm]{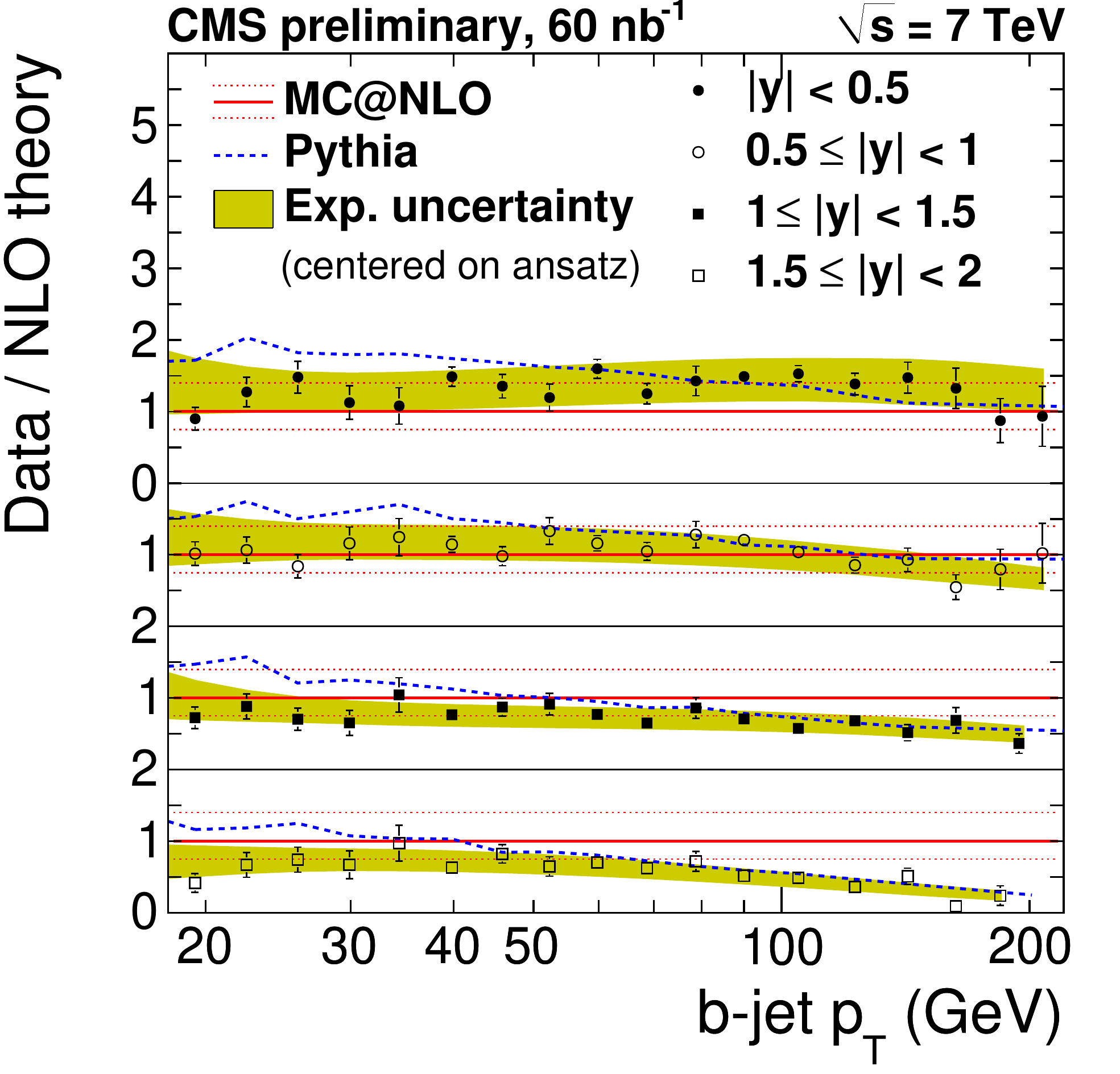}
\caption{B-jets double differential cross section in rapidity and transverse momentum (left) and ratio to theory (right).}
\label{fig:resb}
\end{figure}
The results are shown in Figure \ref{fig:resb} where the differential cross section in rapidity and transverse momentum are shown.

\section{Conclusions}
The first data taken with the CMS detector at $\sqrt{s}=7$~TeV allowed to perform a series of tests and measurement of the capabilities of the CMS detector. Preliminary measurements have been presented both in the exclusive B decay context and in the inclusive B field.

\end{document}